\documentclass[letterpaper, 10 pt, conference]{ieeeconf}  

\IEEEoverridecommandlockouts                              

\usepackage{graphicx} 
\usepackage{amsmath} 

\title{\LARGE \bf
State Estimation and Control for Continuous-Time Nonlinear Systems: A Unified SDRE-Based Approach
}

\author{Azra Redzovic$^{1}$ and Adnan Tahirovic$^{2}$
\thanks{$^{1}$Azra Redzovic is with Faculty of Electrical Engineering, University of Sarajevo, 71000, Bosnia and Herzegovina
        {\tt\small aredzovic1@etf.unsa.ba}}%
\thanks{$^{2}$Adnan Tahirovic is with Faculty of Electrical Engineering, University of Sarajevo, 71000, Bosnia and Herzegovina
        {\tt\small atahirovic@etf.unsa.ba}}%
}

\usepackage{hyperref}

\begin{document}
\maketitle

\vspace{-2ex}
\noindent\textbf{Note:} This is the author’s accepted manuscript of a paper published in the Proceedings of the 2025 IEEE International Conference on Control, Decision and Information Technologies (CoDIT).\\
DOI: \href{https://doi.org/10.1109/CoDIT66093.2025.11321865}{10.1109/CoDIT66093.2025.11321865}

\thispagestyle{empty}
\pagestyle{empty}

\begin{abstract}
This paper introduces a unified approach for state estimation and control of nonlinear dynamic systems, employing the State-Dependent Riccati Equation (SDRE) framework. The proposed approach naturally extends classical linear quadratic Gaussian (LQG) methods into nonlinear scenarios, avoiding linearization by using state-dependent coefficient (SDC) matrices. An SDRE-based Kalman filter (SDRE-KF) is integrated within an SDRE-based control structure, providing a coherent and intuitive strategy for nonlinear system analysis and control design. To evaluate the effectiveness and robustness of the proposed methodology, comparative simulations are conducted on two benchmark nonlinear systems: a simple pendulum and a Van der Pol oscillator. Results demonstrate that the SDRE-KF achieves comparable or superior estimation accuracy compared to traditional methods, including the Extended Kalman Filter (EKF) and the Particle Filter (PF). These findings underline the potential of the unified SDRE-based approach as a viable alternative for nonlinear state estimation and control, providing valuable insights for both educational purposes and practical engineering applications.
\end{abstract}

\section{INTRODUCTION}
\label{introduction}

Accurate state estimation plays a fundamental role in modern control systems, where many control strategies based on state-space description of dynamic systems, including Linear Quadratic Regulator (LQR), pole placement, and Model Predictive Control, rely on precise knowledge of system states to achieve optimal performance. In many practical applications, not all state variables are directly measurable due to sensor limitations, cost constraints, or physical inaccessibility. Consequently, state estimation techniques are essential for reconstructing unmeasured states from available noisy sensor measurements. Traditional estimation methods, such as Kalman estimation, provide a means to infer system states while mitigating the effects of noise and uncertainties, enabling the implementation of advanced control policies in real-world scenarios. However, when dealing with nonlinear systems, conventional state estimation approaches often require approximations or linearization, which can degrade estimation accuracy and, consequently, control performance. This challenge has motivated the development of more advanced estimation techniques that preserve nonlinear system properties, ensuring more reliable state feedback for control applications.

Many real-world control problems involve nonlinear dynamics, including robotics \cite{c17}, traffic management \cite{c18}, spacecraft attitude control \cite{c4}, and biomedical systems such as cancer treatment optimization \cite{c5}, where conventional linear control techniques such as LQR and pole placement often fail to provide satisfactory performance. The State-Dependent Riccati Equation (SDRE) method has gained significant attention due to its conceptual similarity to LQR, making it both intuitive and computationally efficient (see, e.g. \cite{c16} and \cite{c161}). This approach reformulates the nonlinear system into a linear-like structure by expressing its dynamics in terms of state-dependent coefficient (SDC) matrices, allowing optimal control laws to be computed using a Riccati equation at each state. This pointwise factorization enables SDRE-based control to retain nonlinear system properties while leveraging the computational advantages of LQR-like optimal control, making it particularly suitable for complex and highly dynamic systems. Recent developments have extended the SDRE framework by integrating policy iteration methods, enhancing its performance in optimal nonlinear control \cite{c162,c163}. Furthermore, recent studies suggest that SDRE-based policy iteration could be employed as a robust control strategy for nonlinear systems within an integral sliding mode control framework, analogously to previous approaches that robustified the classical LQR for linear systems \cite{c164}.

Reliable state estimation for nonlinear dynamic systems is crucial in modern engineering and control applications. 
Traditional techniques, like the Extended Kalman Filter (EKF), simplify the problem by approximating nonlinear systems as linear around their operating points. While this simplification makes the method easier to apply, it can lead to reduced accuracy, especially when the system experiences strong nonlinearities or disturbances. Particle filters (PF), on the other hand, handle nonlinearities without approximations, but they typically require significant computational resources, making them difficult to implement in real-time scenarios. In contrast, the SDRE-based Kalman filter (SDRE-KF) provides a practical solution by directly using the system’s nonlinear dynamics, maintaining high accuracy and requiring lower computational effort than in case of the PF. 

Recent studies have explored nonlinear state estimation tehniques, emphasizing improved control performance and estimation accuracy. For instance, a novel sampled-data non-affine nonlinear observer has been presented for state estimation of nonlinear networked systems subject to aperiodic sampled delayed measurements, outperforms existing methods in terms of convergence speed and estimation accuracy \cite{ref1}. Finite- and fixed-time parameter estimation and control method has been proposed for continuous-time nonlinear systems, achieving the convergence of system states even when the estimation is not accurate \cite{ref2}. Application of SDRE-based control methods combined with nonlinear state estimation techniques has been increasingly explored in recent studies. A finite-time SDRE controller using online update of the state-dependent coefficients has been successfully utilized for anthropomorphic dual-arm space manipulator system in free-flying conditions, outperforming LQR in terms of better positioning errors and showing efficacy of the proposed method \cite{ref3}. Evaluation of the suitability of SDRE control for hovering of a spacecraft in the highly nonlinear environment demonstrated flexibility and effectiveness, obtaining the absolute and relative errors of the order of $10^{-14}$ \cite{ref4}. An experimental study on the quadrotor has indicated  promising implementation and success in terms of tracking \cite{ref5}.

The main contribution of this paper is the unification of the SDRE approach for both control and state estimation in nonlinear dynamic systems. Specifically, we introduce and evaluate a unified framework that combines SDRE-based control with SDRE-based Kalman filtering, presenting it as a natural nonlinear extension of the well-known LQG control structure, which pairs LQR control with Kalman filtering for linear systems. Such a unified SDRE-based framework offers significant educational and practical value, as it provides a systematic and intuitive approach to transitioning from linear control and estimation techniques toward more complex nonlinear methods. The second contribution lies in demonstrating how the SDRE methodology avoids oversimplification inherent to traditional methods like linearization, ensuring the retention of critical nonlinear dynamics. Finally, the third key contribution of this work is a comprehensive comparative analysis between the SDRE-KF, the EKF, and PF within the context of SDRE-based nonlinear control, providing practical insights into their relative strengths, limitations, and suitability for real-time implementation.

Section~\ref{control} introduces the SDRE-based control approach for continuous-time nonlinear systems, detailing its formulation and implementation. Section~\ref{estimation} describes the SDRE-KF for state estimation and discusses standard nonlinear estimation methods used for comparative analysis, specifically the EKF and PF. A thorough performance comparison, including simulation results, is presented in Section~\ref{simulation}, with an emphasis on accuracy under the SDRE-based control framework. Finally, Section~\ref{conclusion} summarizes key findings and outlines future research directions.

\section{SDRE BASED CONTROL FOR CONTINUOUS-TIME SYSTEMS}
\label{control}
The SDRE provides a practical and effective framework for synthesizing nonlinear feedback controllers (see, e.g. \cite{c162}, \cite{c163}, \cite{c13}, \cite{c1}). The key idea behind the SDRE approach is to factorize the original nonlinear system dynamics, representing the nonlinear system in a linear-like form through SDC matrices:
\begin{equation}
    \dot{x} = A(x)x + B(x)u
\end{equation}
where \textit{x} denotes the system states, \textit{A} is the state matrix, 
\textit{B} is the input matrix, and \textit{u} represents the control input.

The control objective typically involves minimizing an infinite-horizon, quadratic-like performance criterion of the form:
\begin{equation}
    J = \frac{1}{2}\int_{0}^{\infty}\left(x^TQ(x)x + u^TR(x)u\right)dt,
\end{equation}
where the weighting matrices $Q(x)$ and $R(x)$ satisfy $Q(x)=D^T(x)D(x)\geq0$ and $R(x)>0$ for all states $x$ \cite{c16}.

Under these conditions, the state-feedback control law can be expressed as
\begin{equation}
u(x) = -K(x)x = -R^{-1}(x)B^{T}(x)P(x)x,
\end{equation}
where the matrix $P(x)$ is the symmetric, positive-definite solution to the state-dependent algebraic Riccati equation given by \cite{c16}:
\begin{equation}
\begin{aligned}
P(x)A(x)& + A^T(x)P(x)  \\
- &P(x)B(x)R^{-1}(x)B^T(x)P(x) + Q(x) = 0
\end{aligned}
\end{equation}
The solution of this optimization problem requires that the system be controllable, which can be verified by checking the rank of the state-dependent controllability matrix, as given by \cite{c15}:
\begin{equation}
rank
\left[B(x) \ A(x)B(x) \ \cdots \ A(x)^{n-1}B(x)\right] = n,
\end{equation}
where \textit{n} denotes the dimension of the state vector.

A key aspect of the SDRE approach is the nonuniqueness of the SDC parameterization for multivariable systems, which provides additional degrees of freedom. Specifically, given two distinct parameterizations $A_1(x)$ and $A_2(x)$  where $f(x)=A_1(x)x=A_2(x)x$, any convex combination defined by 
\begin{equation}
A(x,\alpha) = \alpha A_1(x) + (1-\alpha)A_2(x)
\end{equation}
is also a valid SDC parameterization \cite{c1}.

SDRE control offers a similar approach as an algebraic Riccati equation (ARE) for a LQR \cite{c1}. However, since SDRE is not derived from the Hamilton–Jacobi–Bellman equation, it does not provide an optimal solution when it comes to the optimal control of the nonlinear system \cite{c162}, \cite{c163}.

\section{NONLINEAR STATE ESTIMATION}
\label{estimation}

\subsection{Extended Kalman filter}
The EKF estimates states of nonlinear systems by linearizing system dynamics around the current estimated operating point (see, e.g. \cite{c6}). Although widely adopted in practice due to its ability to yield good estimation performance, the EKF may suffer from poor accuracy and can even become unstable when applied to highly nonlinear systems. This limitation significantly differentiates it from the standard linear Kalman filter, which typically exhibits consistent stability and reliable performance under linear conditions (see, e.g. \cite{c7}). Additionally, unlike the linear Kalman filter, the EKF provides only an approximate rather than optimal state estimate due to the inherent linearization step involved (see, e.g. \cite{c8}).

Consider a general nonlinear continuous-time system described by the following state-space representation:
\begin{equation}
    \begin{aligned}
        \dot{x}(t) &= f\left(x(t),u(t),w(t),t\right)\\
        y(t) &= h\left(x(t),v(t),t\right)\\
        w(t) &\sim \mathcal{N}(0,Q)\\
        v(t) &\sim \mathcal{N}(0,R)
    \end{aligned}
\end{equation}
where \( x(t) \) represents the state vector, \( u(t) \) is the input vector, and \( y(t) \) is the measurement vector. The process noise \( w(t) \) and measurement noise \( v(t) \) are assumed Gaussian with zero mean and covariance matrices \( Q \) and \( R \), respectively.

The EKF addresses the nonlinear state estimation problem by linearizing the nonlinear system around the current estimated state. This linearization involves computing partial derivatives of the system and measurement functions evaluated at the current estimated state as follows \cite{c6}:
\begin{equation}
    A = \left.\frac{\partial f}{\partial x}\right|_{\hat{x}},\quad
    L = \left.\frac{\partial f}{\partial w}\right|_{\hat{x}},\quad
    C = \left.\frac{\partial h}{\partial x}\right|_{\hat{x}},\quad
    M = \left.\frac{\partial h}{\partial v}\right|_{\hat{x}}.
\end{equation}
Using these matrices, the effective process noise covariance matrix $\Tilde{Q}$ and the measurement noise covariance matrix $\Tilde{R}$ are calculated as \cite{c6}:
\begin{equation}
    \Tilde{Q} = LQL^T, \quad \Tilde{R} = MRM^T.
\end{equation}

The EKF algorithm can then be summarized by the following equations:
\begin{equation}
\begin{aligned}
\dot{\hat{x}}(t) &= f(\hat{x}(t),u(t),0,t) + K(t)\left(y(t)-h(\hat{x}(t),0,t)\right),\\[6pt]
K(t) &= P(t)C^T(t)\Tilde{R}^{-1}(t),\\[6pt]
\dot{P}(t) &= A(t)P(t) + P(t)A^T(t) - K(t)\Tilde{R}(t)K^T(t) + \Tilde{Q}(t),
\end{aligned}
\end{equation}
with initial conditions given by:
\begin{equation}
\hat{x}(0) = E[x(0)],\quad P(0) = E\left[(x(0)-\hat{x}(0))(x(0)-\hat{x}(0))^T\right].
\end{equation}

\subsection{Particle filter}
The PF is a probabilistic state estimation method suitable for highly nonlinear systems (see, e.g. \cite{c6}). Although suboptimal in the strict sense \cite{c9}, it provides effective numerical approximations of complex estimation problems using adaptive stochastic grids that dynamically select relevant state points, with linear complexity regarding the number of points selected \cite{c10}. Particle filters are widely employed in robotics, particularly in problems such as robot localization, due to their flexibility in handling nonlinearities and uncertainties \cite{c11}. However, achieving higher estimation accuracy with particle filters typically requires increased computational resources, which may limit their real-time applicability depending on the complexity of the particular problem at hand \cite{c6}.

Consider the following general discrete-time nonlinear system:
\begin{equation}
        x_{k+1} = f(x_k,u_k,w_k), \ y_k = h(x_k,u_k,v_k),
\end{equation}
where the process noise $w_k$ and measurement noise $v_k$ are independent white-noise processes characterized by known probability density functions (pdfs).

The PF algorithm proceeds as follows. Initially, a set of $N$ random state vectors, called particles, is generated based on the pdf, $p(x_0)$, of the initial state, and denoted as $x_{0,i}^{+},\;i=1,...,N$. Subsequently, for each time step $k=1,2,\dots$, the following iterative steps are performed \cite{c6}:
\begin{itemize}
    \item \textbf{Prediction step}: Compute the predicted (\textit{a priori}) particles using the process model:
    \begin{equation}
        x_{k,i}^{-} = f(x_{k-1,i}^{+},u_{k-1},w_{k-1,i}).
    \end{equation}

    \item \textbf{Update step}: Using the current measurement $y_k$, evaluate the relative likelihood $q_i$ of each particle according to the measurement pdf $p(y_k|x_{k,i}^{-})$.

    \item \textbf{Normalization}: Normalize the likelihoods $q_i$ such that their sum equals one, forming a probability distribution over particles.

    \item \textbf{Resampling step}: Draw a new set of \textit{a posteriori} particles $x_{k,i}^{+}$ according to their normalized likelihoods. This step is crucial as it prevents particle degeneration and ensures filter stability over time \cite{c10}.

    \item \textbf{Estimation}: Compute the statistical characterization of the posterior distribution $p(x_k|y_k)$, typically represented by the mean and covariance of the resampled particles.
\end{itemize}

\subsection{SDRE-based Kalman filter}
The SDRE method has been successfully extended to address state estimation problems for nonlinear systems. In particular, an SDRE-based estimator offers a natural nonlinear extension of the standard Kalman filtering framework. Specifically, while both SDRE-based control and estimation reduce to conventional linear control and estimation (such as LQR and standard Kalman filtering, respectively) for linear systems or constant system matrices, their state-dependent structure provides significant advantages when dealing with nonlinear dynamics. Unlike traditional estimation methods that require linearization, the SDRE-KF directly leverages SDC matrices, thereby accurately capturing nonlinear system dynamics and improving estimation robustness and accuracy.

Consider a general continuous-time nonlinear system represented by:
\begin{equation}
\begin{aligned}
    \dot{x}(t) &= A(x)x(t) + B(x)u(t) + w(t), \\
    y(t) &= C(x)x(t) + v(t),
\end{aligned}
\end{equation}
where \(w(t)\) and \(v(t)\) represent process and measurement noise with covariance matrices \(Q(x)\) and \(R(x)\), respectively. These noise vectors are assumed to be zero-mean Gaussian processes:
\[
w(t)\sim\mathcal{N}(0,Q(x)),\quad v(t)\sim\mathcal{N}(0,R(x)).
\]
When system matrices \(A(x)\), \(B(x)\), and \(C(x)\) become constant (state-independent), the SDRE-based Kalman estimator simplifies to the well-known linear Kalman filter. For the general nonlinear scenario, the SDRE-based filter equations are given by:
\begin{equation}
\begin{aligned}
    \dot{\hat{x}}(t) &= A(x)\hat{x}(t) + B(x)u(t) + K_f(x)\left[y(t)-C(x)\hat{x}(t)\right],\\[6pt]
    K_f(x) &= P(x)C^T(x)R^{-1}(x),\\[6pt]
    0 &= P(x)A^T(x) + A(x)P(x) \\
    & - P(x)C^T(x)R^{-1}(x)C(x)P(x) + Q(x).
\end{aligned}
\end{equation}

A necessary condition for successful filter convergence is that the nonlinear system remains completely observable at each time step. This state-dependent observability condition is verified by ensuring the state-dependent observability matrix:
\begin{equation}
\text{rank}
\begin{bmatrix}
C^T(x)\ (C(x)A(x))^T  \dots  (C(x)A^{n-1}(x))^T
\end{bmatrix}^T = n
\end{equation}
where \(n\) denotes the number of system states \cite{c14}.

\section{SIMULATION RESULTS}
\label{simulation}
\subsection{Simple pendulum}
A simple pendulum system consists of a mass (\textit{bob}) attached to a rigid, massless rod, pivoted at one end to allow rotation (see, e.g. \cite{c2}). The angle between the rod and the vertical axis is denoted by $\theta$, and the externally applied torque to control the
motion of the pendulum is denoted as $T$. The dynamics of the pendulum are described by the nonlinear differential equation \cite{c2}:
\begin{equation}
    m l \ddot{\theta} = -m g \sin{\theta} - k l \dot{\theta} + \frac{1}{l} T.
\end{equation}
Using the SDRE framework, the system can be represented in a state-dependent linear-like form as:
\begin{equation}
    \begin{bmatrix}
        \dot{\theta}\\[4pt]
        \ddot{\theta}
    \end{bmatrix}
    =
    \begin{bmatrix}
        0 & 1\\[4pt]
        -\frac{g}{l}\frac{\sin\theta}{\theta} & -\frac{k}{m}
    \end{bmatrix}
    \begin{bmatrix}
        \theta\\[4pt]
        \dot{\theta}
    \end{bmatrix}
    +
    \begin{bmatrix}
        0\\[4pt]
        \frac{1}{ml^2}
    \end{bmatrix} T.
\end{equation}

The system parameters used in this study are: pendulum rod length \( l = 1.5\,\text{m} \), bob mass \( m = 0.5\,\text{kg} \), and friction coefficient \( k = 0.5\). The weighting matrices for SDRE-based control and estimation are defined as \( Q = \begin{bmatrix}10 & 0 \\ 0 & 10\end{bmatrix} \) and \( R = 0.1 \), respectively. Additionally, the process disturbance and measurement noise covariance matrices are selected as \( Q_{\text{disturbance}} = \begin{bmatrix}0.1 & 0 \\ 0 & 0.1\end{bmatrix} \) and \( R_{\text{noise}} = 0.1 \). The weighting matrices employed in the SDRE approach were tuned empirically to achieve desirable system performance. Both measurement and process noise are modeled as zero-mean Gaussian. For the PF simulations, the number of particles used is \( N_{\text{particles}} = 500 \). It is assumed that only the angular position \(\theta\) is measured. Hence, the measurement matrix is \( C = [1\;\;0] \). A total of 30 simulations are performed with a sampling time of \(0.01\,\text{s}\). Note that for SDRE-based KF and the EKF, the continuous-time system representation is directly employed, whereas the PF simulations require discretization of the system dynamics.

\begin{figure}[thpb]
  \begin{minipage}{0.49\linewidth}
    \centering
    \includegraphics[width=1\linewidth]{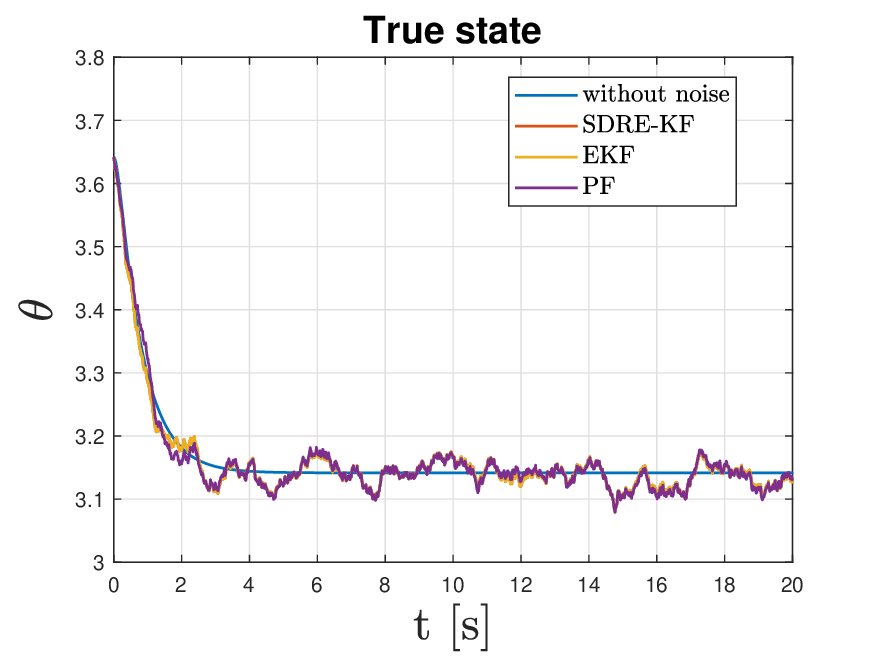}
  \end{minipage}
  \begin{minipage}{0.49\linewidth}
    \centering
    \includegraphics[width=1\linewidth]{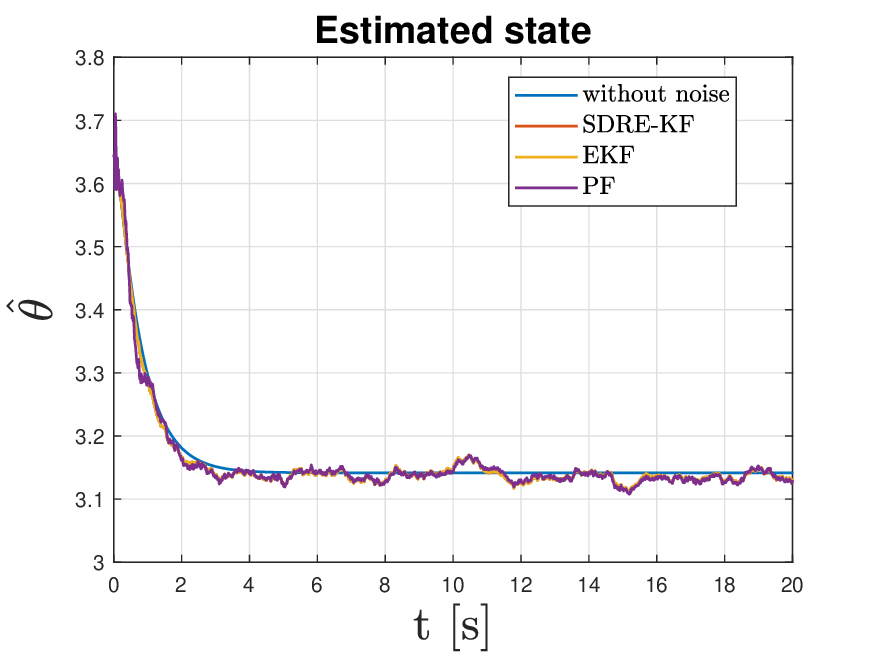}
  \end{minipage}
  \caption{True and estimated state of $\theta$}
  \label{fig_sp1}
\end{figure}

\begin{figure}[thpb]
  \begin{minipage}{0.49\linewidth}
    \centering
    \includegraphics[width=1\linewidth]{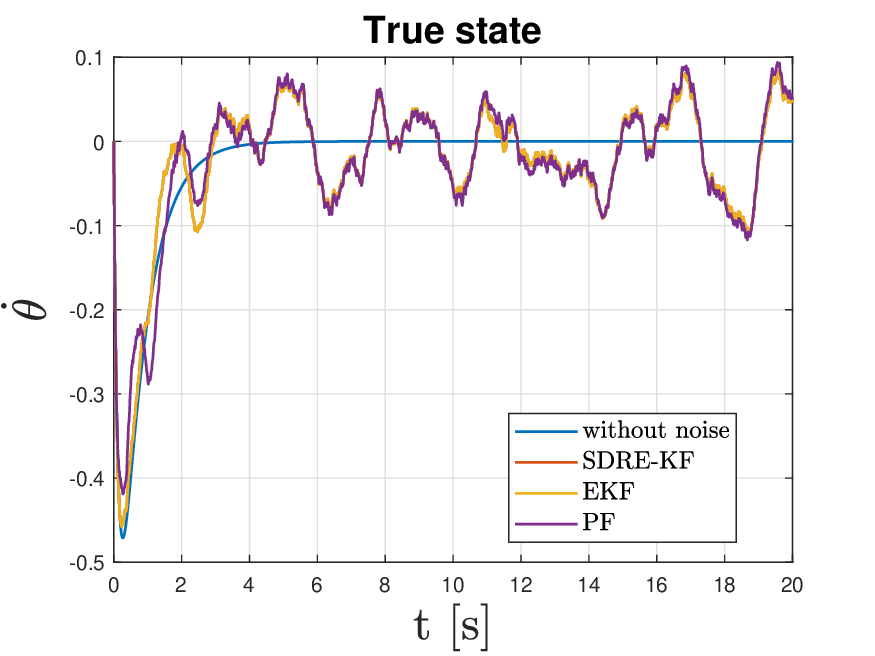}
  \end{minipage}
  \begin{minipage}{0.49\linewidth}
    \centering
    \includegraphics[width=1\linewidth]{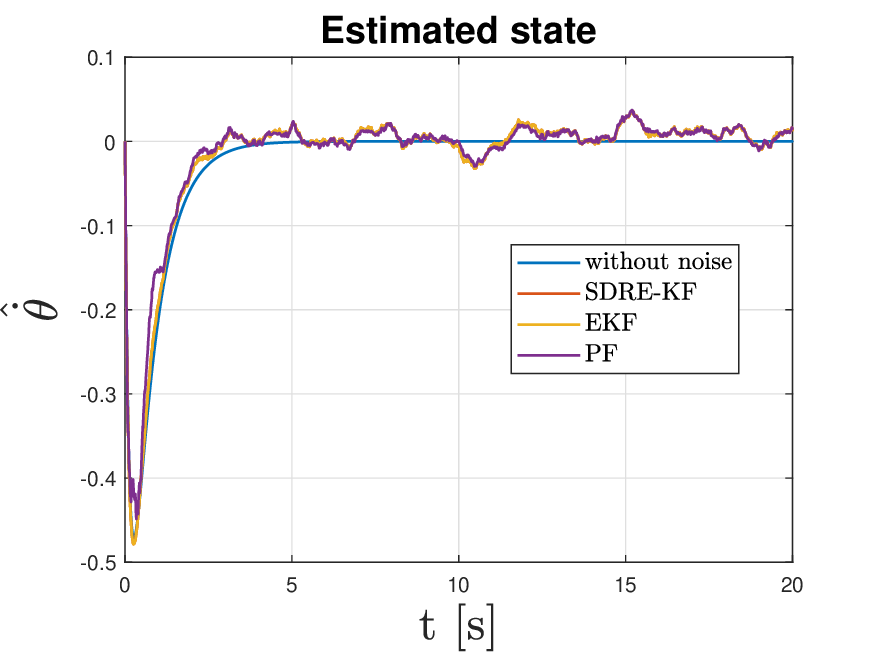}
  \end{minipage}
  \caption{True and estimated state of $\dot{\theta}$}
  \label{fig_sp2}
\end{figure}

\begin{figure}[thpb]
     \centering
   \includegraphics[scale=0.4]{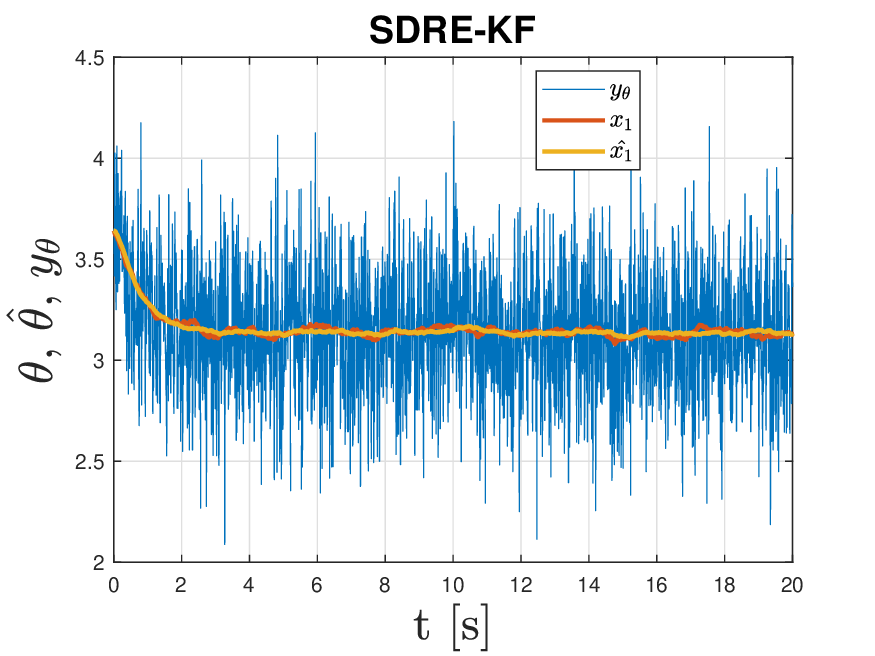}
  \caption{True and estimated state, $\theta$ and $\hat{\theta}$, with noisy measurement, $y_\theta$}
 \label{fig_sp3}
\end{figure}

Simulation results for the simple pendulum system are illustrated in Figs.~\ref{fig_sp1}-\ref{fig_sp3}. Specifically, Fig.~\ref{fig_sp1} compares the true state trajectory with state estimates obtained using the SDRE-KF, EKF, and PF for measured state variable $\theta$. The trajectories illustrate system behavior starting from the initial condition $[\theta, \dot{\theta}]^T = [\pi + 0.5,\, 0]^T$ and converging towards the equilibrium state $[\pi, 0]^T$. Fig.~\ref{fig_sp2} additionally depicts the state variable $\dot{\theta}$, alongside its corresponding estimates provided by the filters. To clearly show the effect of measurement noise and filtering performance, Fig.~\ref{fig_sp3} presents a direct comparison of the noisy measured signal and the corresponding estimated signals. These visualizations demonstrate the effectiveness of SDRE-KF in accurately estimating system states under noisy conditions, especially when paired with SDRE-based control.

Performance evaluation based on statistical metrics across 30 independent simulations is summarized in Tables~\ref{tab_sp_mse} and~\ref{tab_sp_mae}. The Mean Squared Error (MSE), and the Mean Absolute Error (MAE) are computed for each estimation method, namely the SDRE-KF, EKF, and PF, using the simple pendulum model. The presented results clearly indicate that the SDRE-KF filter consistently outperforms the other methods, achieving the lowest values of MSE and MAE. These results confirm the advantage of using the SDRE-based filtering approach, particularly in terms of accuracy and robustness for state estimation of nonlinear dynamic systems.

\begin{table}[thpb]
\caption{MSE comparison of SDRE-KF, EKF and PF by states for simple pendulum}
\label{tab_sp_mse}
\begin{center}
\begin{tabular}{|c|c|c|c|}
\hline
& \multicolumn{3}{c|}{\textbf{MSE}} \\
\hline
 & \textbf{SDRE-KF} & \textbf{EKF} & \textbf{PF} \\
\hline
$\theta$ & \textbf{0.00048} & 0.00051 & 0.00067 \\
\hline
$\dot{\theta}$ & \textbf{0.00251} & 0.00257 & 0.00324 \\
\hline
\end{tabular}
\end{center}
\end{table}


\begin{table}[thpb]
\caption{MAE comparison of SDRE-KF, EKF and PF by states for simple pendulum}
\label{tab_sp_mae}
\begin{center}
\begin{tabular}{|c|c|c|c|}
\hline
& \multicolumn{3}{c|}{\textbf{MAE}} \\
\hline
 & \textbf{SDRE-KF} & \textbf{EKF} & \textbf{PF} \\
\hline
$\theta$ & \textbf{0.01760} & 0.01786 & 0.01982 \\
\hline
$\dot{\theta}$ & \textbf{0.03943} & 0.03991 & 0.04396 \\
\hline
\end{tabular}
\end{center}
\end{table}

\subsection{Van der Pol oscillator}
Consider the Van der Pol oscillator described by the following second-order nonlinear differential equation \cite{c162}:
\begin{equation}
    \dot{x_1} = x_2,\ \dot{x_2} = -x_1 - \mu(1-x_1^2)x_2 + x_1u
\end{equation}
The Van der Pol oscillator dynamics can be factorized into the following state-dependent linear-like representation:
\begin{equation}
    \begin{bmatrix}
        \dot{x}_1 \\[4pt]
        \dot{x}_2
    \end{bmatrix}
    =
    \begin{bmatrix}
        0 & 1\\[4pt]
        -1 & -\mu(1 - x_1^2)
    \end{bmatrix}
    \begin{bmatrix}
        x_1 \\[4pt]
        x_2
    \end{bmatrix}
    +
    \begin{bmatrix}
        0 \\[4pt]
        x_1
    \end{bmatrix}u.
\end{equation}

The parameters used for the system simulation include the nonlinearity coefficient \(\mu = 0.7\), weighting matrices \(Q = \begin{bmatrix}1 & 0\\0 & 1\end{bmatrix}\), \(R = 0.1\), process disturbance covariance matrix \(Q_{\text{disturbance}} = \begin{bmatrix}0.1 & 0\\0 & 0.1\end{bmatrix}\), measurement noise covariance \(R_{\text{noise}} = 0.1\), and the number of particles for the PF, \(N_{\text{particles}} = 500\). Both measurement and process noise are modeled as zero-mean Gaussian. The weighting matrices used in the SDRE-based control and estimation design were adjusted based on observed system performance. Only the state variable \(x_1\) is assumed to be directly measured, yielding a measurement matrix \(C = [1\quad 0]\). A simulation time step of \(0.01\,\text{s}\) is adopted. Continuous-time dynamics are directly employed in SDRE-KF and EKF, while discretization is performed for PF implementation.

\begin{figure}[thpb]
  \begin{minipage}{0.49\linewidth}
    \centering
    \includegraphics[width=1\linewidth]{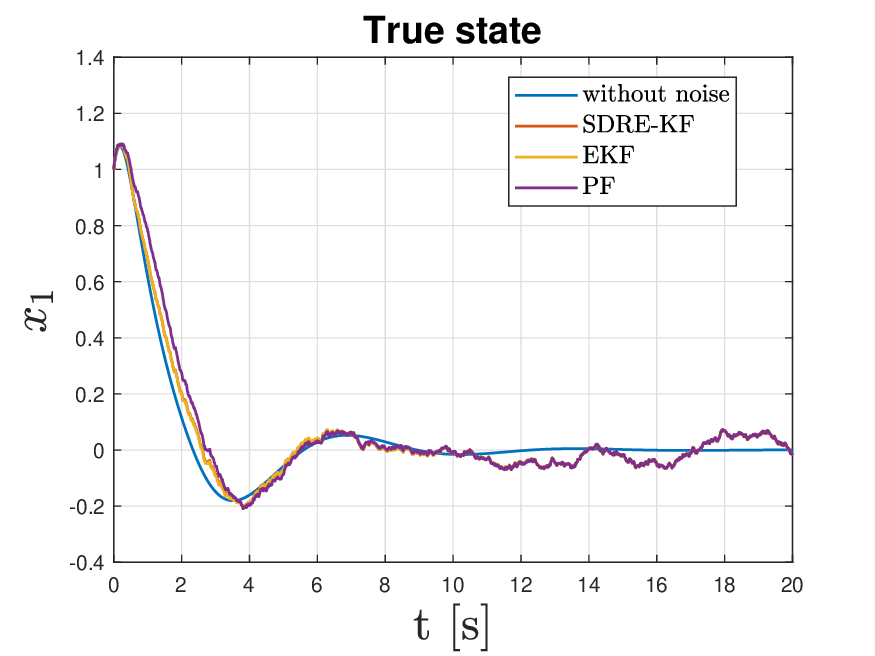}
  \end{minipage}
  \begin{minipage}{0.49\linewidth}
    \centering
    \includegraphics[width=1\linewidth]{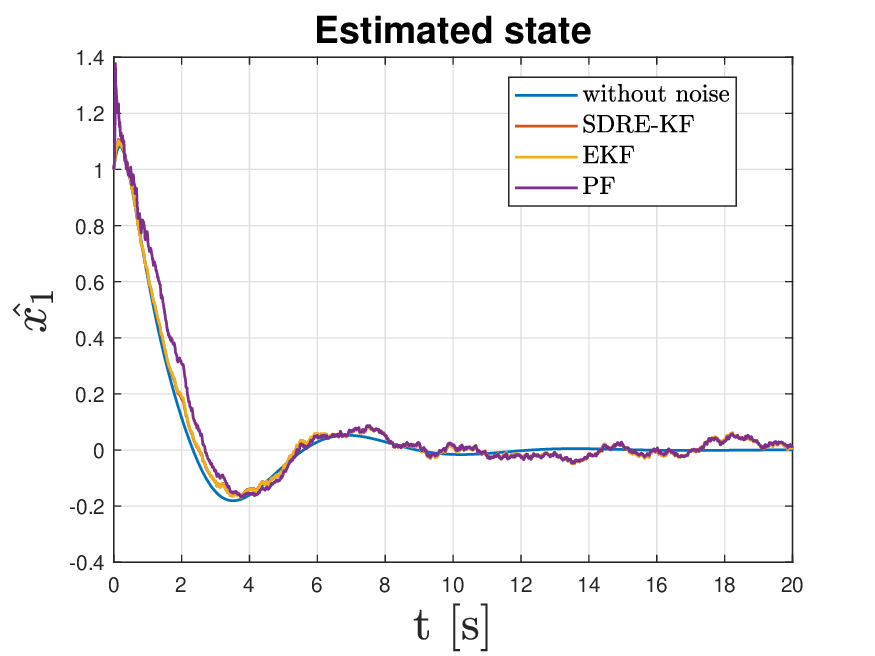}
  \end{minipage}
  \caption{True and estimated state of $x_1$}
  \label{fig_vdp1}
\end{figure}

\begin{figure}[thpb]
  \begin{minipage}{0.49\linewidth}
    \centering
    \includegraphics[width=1\linewidth]{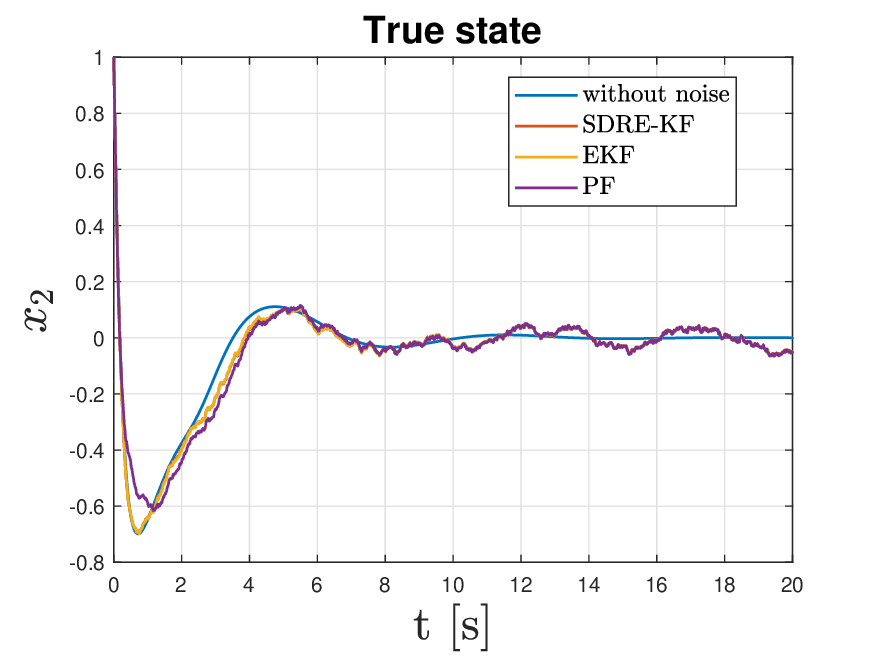}
  \end{minipage}
  \begin{minipage}{0.49\linewidth}
    \centering
    \includegraphics[width=1\linewidth]{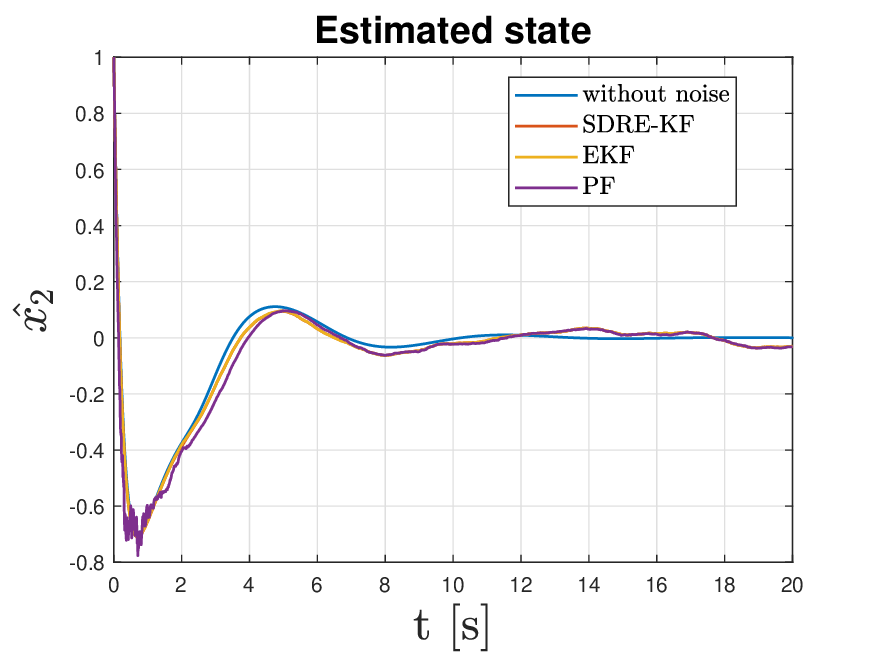}
  \end{minipage}
  \caption{True and estimated state of $x_2$}
  \label{fig_vdp2}
\end{figure}

\begin{figure}[thpb]
     \centering
   \includegraphics[scale=0.4]{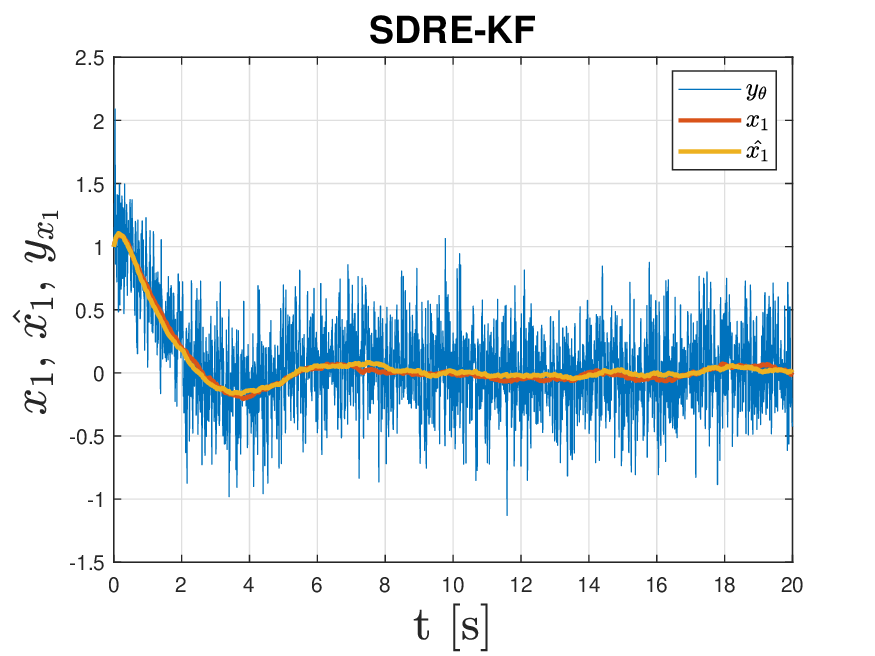}
  \caption{True and estimated state, $x_1$ and $\hat{x_1}$, with noisy measurement, $y_{x_1}$}
 \label{fig_vdp3}
\end{figure}

The true and estimated states of the Van der Pol oscillator, using the SDRE-KF, EKF, and PF, are illustrated in Fig.~\ref{fig_vdp1} (state \(x_1\)) and Fig.~\ref{fig_vdp2} (state \(x_2\)). Additionally, these figures include the reference trajectory, obtained from system simulations employing only SDRE-based control without estimation. Fig.~\ref{fig_vdp3} presents the measured noisy signal along with the corresponding estimated state trajectories. The presented signals illustrate the system response from an initial condition of \([1;\,1]\) toward the equilibrium state \([0;\,0]\).

Finally, Tables~\ref{tab_vdp_mse} and~\ref{tab_vdp_mae} summarize a statistical comparison between the SDRE-KF, EKF, and PF methods based on 30 independent simulations, using metrics such as MSE and MAE. The results indicate that for the state \(x_1\), the PF achieves the lowest estimation errors, whereas for state \(x_2\), the EKF demonstrates superior performance. Although the SDRE-KF did not achieve the lowest errors in this particular scenario, its performance remained competitive, producing estimation results closely comparable to the best-performing methods.

Although explicit comparison of execution time and resource usage was not included in this study, it is worth noting that EKF and SDRE-KF generally have similar computational demands, while PF typically requires more resources due to the large number of particles.

\begin{table}[thpb]
\caption{MSE comparison of SDRE-KF, EKF and PF by states for Van Der Pol oscillator}
\label{tab_vdp_mse}
\begin{center}
\begin{tabular}{|c|c|c|c|}
\hline
& \multicolumn{3}{c|}{\textbf{MSE}} \\
\hline
 & \textbf{SDRE-KF} & \textbf{EKF} & \textbf{PF} \\
\hline
$x_1$ & 0.00334 & 0.00231 & \textbf{0.00151} \\
\hline
$x_2$ & 0.01379 & \textbf{0.01160} & 0.01189 \\
\hline
\end{tabular}
\end{center}
\end{table}


\begin{table}[thpb]
\caption{MAE comparison of SDRE-KF, EKF and PF by states for Van Der Pol oscillator}
\label{tab_vdp_mae}
\begin{center}
\begin{tabular}{|c|c|c|c|}
\hline
& \multicolumn{3}{c|}{\textbf{MAE}} \\
\hline
 & \textbf{SDRE-KF} & \textbf{EKF} & \textbf{PF} \\
\hline
$x_1$ & 0.03255 & 0.03005 & \textbf{0.02821} \\
\hline
$x_2$ & 0.04001 & \textbf{0.03715} & 0.03770 \\
\hline
\end{tabular}
\end{center}
\end{table}

\section{CONCLUSIONS}
\label{conclusion}
This paper presented a unified approach for addressing nonlinear state estimation by combining SDRE-based control with an SDRE-KF. The SDRE framework naturally extends the classical LQG methodology to nonlinear systems without resorting to linearization. The advantage of this unified approach lies in its intuitive appeal, particularly for educational purposes, as it provides a consistent strategy for both control and estimation in nonlinear system analysis and design.

Through comprehensive simulations involving benchmark nonlinear systems, including the simple pendulum and the Van der Pol oscillator, we evaluated the SDRE-KF against conventional nonlinear filters such as the EKF and PF. Results confirmed that the SDRE-KF method can achieve comparable or superior estimation accuracy and robustness, depending on the system's specific nonlinear characteristics. Specifically, SDRE-KF outperformed EKF and PF in one of the studied systems by providing minimal estimation errors. In the second scenario, although the SDRE-KF did not achieve the best results, its performance remained competitive and closely comparable to the best-performing method.

The findings of this study emphasize the potential of the SDRE-based estimation approach as a viable alternative to traditional nonlinear estimation techniques. Future work will explore adaptive strategies for selecting optimal state-dependent parameterizations, aiming to further enhance estimation accuracy across a wider range of nonlinear dynamic systems. Future research will also explore novel SDRE-based strategies utilizing the policy iteration paradigm to achieve optimal nonlinear state estimation, following the approaches proposed in \cite{c162,c163}.

\addtolength{\textheight}{-12cm}   



\section*{Acknowledgment}
This work has been supported in part by the scientific project "Strengthening Research and Innovation Excellence in Autonomous Aerial Systems - AeroSTREAM," supported by the European Commission HORIZON WIDERA-2021-ACCESS-05 Programme through the project under G.A. number 101071270.

\bibliographystyle{unsrt}

\end{document}